\documentclass{kluwer}
\usepackage{graphicx}

\newdisplay{guess}{Conjecture}

\begin{document}
\begin{article}
\begin{opening}
\title{On the Saturation of Astrophysical Dynamos:
Numerical Experiments with the No-cosines flow}

\author{S.B.F. \surname{Dorch}}

\runningauthor{Dorch and Archontis} \runningtitle{The no-cosines
dynamo}

\institute{The Niels Bohr Institute for Astronomy, Physics and
Geophysics, Juliane Maries Vej 30, DK-2100 Copenhagen {\O},
Denmark}

\author{V. \surname{Archontis}}
\institute{Instituto de Astrofisica de Canarias, Via Lactea s/n
E-38200, La Laguna, Spain}

\date{August 16, 2004}

\begin{abstract}
In the context of astrophysical dynamos we illustrate that the
no-cosines flow, with zero mean helicity, can drive fast dynamo
action and study the dynamo's mode of operation during both the
linear and non-linear saturation regime: It turns out that in
addition to a high growth rate in the linear regime, the dynamo
saturates at a level significantly higher than normal turbulent
dynamos, namely at exact equipartition when the magnetic Prandtl
number Pr$_{\rm m} \sim 1$. Visualization of the magnetic and
velocity fields at saturation will help us to understand some of
the aspects of the non-linear dynamo problem.
\end{abstract}

\end{opening}

\section{Introduction}

Magnetic fields are common in astronomical objects and are
responsible for a variety of complex phenomena in the Universe: It
is widely accepted that these fields are generated by the motions
of conducting fluids through a transfer of kinetic to magnetic
energy, that is, dynamo action \cite{Parker1979}. However, our
understanding of the exact mechanism of generation and maintenance
of the magnetic field against dissipation is still incomplete even
for simple dynamos. On the observational side, when a magnetic
field in an astronomical object is not observed directly often the
inferred equipartition value $B_{\rm eq} = u\sqrt{\mu_0\rho}$
(where $u$ and $\rho$ are characteristic speed and mass density,
and $\mu_0$ is the vacuum permeability) is assumed, e.g.\
\inlinecite{Safier1999}: Therefore it is of importance to consider
whether this is a good approach. In the Sun the equipartition
field strength at the surface is about 500 Gauss. However, the
mean magnetic field strength near the Sun's polar regions has a
canonical value of 1--5 G going back to Babcock, which is far less
than $B_{\rm eq}$; additionally, the intermittent flux tubes at
the surface is much stronger, with field strengths around 1--3 kG
in spots and flux tubes, while it is inferred to be 10 kG near the
bottom of the convection zone, e.g.\ \inlinecite{Moreno+ea1994}.
In any case $B_{\rm eq}$ is not a good estimate for the solar
magnetic field. Furthermore, on the theoretical side of things,
numerical models of turbulent non-linear dynamo action as a rule
produce strong intermittent structures, but cannot be said to
generally saturate at equipartition; these models can be used to
study the generic properties of real astrophysical dynamos, such
as the solar dynamo \cite{Choudhuri2003}.

Dynamo action in the limit of infinite magnetic Reynolds number
${\rm Re}_{\rm m}=u\ell/\eta$ (where $\ell$ is a characteristic
length scale and $\eta$ is the magnetic diffusivity) is relevant
to astrophysical systems where the diffusion time-scale is larger
than the advection time-scale, such as in the Sun. Considering the
importance of the magnetic forces relative to the motion of the
fluid, one divides dynamo action into two regimes: The linear
kinematic regime in which the flow amplifies the magnetic field
exponentially by e.g.\ stretching the magnetic field lines
\cite{STF} and the non-linear regime where the magnetic field
becomes strong enough to modify the initial flow topology through
the Lorentz force and saturate the exponential growth.

Recent studies \cite{Dorch2000,Archontis+ea03a,Tanner+Hughes2003}
have improved our knowledge of fundamental dynamo mechanisms of
prescribed flows in the kinematic regime. Non-linear fast dynamo
action has received considerable attention and progress has been
made especially by means of numerical magneto-hydrodynamical (MHD)
simulations \cite{Nordlund+ea92,Cattaneo+ea96}. However, it is
still unclear what are the physical processes at work before and
especially after the saturation of the dynamo: It is generally
found that dynamos saturate because of a suppression of the
stretching ability of the flows
\cite{Cattaneo+ea96,Archontis+ea03,Tanner+Hughes2003}, but the
details of how this comes about are not known.

In this paper, we present results from numerical experiments with
a non-helical flow first studied by
\inlinecite{Galloway+Proctor1992} that saturates below
equipartition only for high values of the Reynolds number ${\rm
Re}=u\ell/\nu$ ($\nu$ being the viscosity): At magnetic Prandtl
number ${\rm Pr}_{\rm m}={\rm Re}_{\rm m}/{\rm Re} \sim 1$
saturation occurs at almost exact equipartition. We study the
non-linear dynamo mode of operation, examine how it changes when
the flow becomes turbulent, and take the first steps at
illustrating how the fields evolve and interact.
\section{Setup of the numerical experiments}

The fully compressible 3-d MHD equations are solved numerically on
a periodic Cartesian grid:
\begin{eqnarray}
{\frac{\partial {\bf \rho}} {\partial t}} & = & -\nabla\cdot\rho{\bf u}, \label{mass.eq}\\
{\frac{\partial {({\rho}{\bf u})}} {\partial t}}& = & -\nabla P
    + {\bf j}\times{\bf B} + {\bf f} - \nabla\cdot(\rho{\bf u}{\bf u}) \label{motion.eq},\\
{\frac{\partial e}  {\partial t}} & = & - \nabla\cdot(e{\bf u})
    - P\nabla\cdot{\bf u} + {\rm Q}_{\rm v} + {\rm Q}_{\rm J} + {\rm Q}_{\rm cool},
    \\ \label{energy.eq}
\frac{\partial {\bf B}}{\partial t} & =  &\nabla \times ( {\bf u}
\times {\bf B})
    + \frac{1}{{\rm Re}_{\rm m}} \nabla^2 {\bf B}, \label{induction.eq}
\end{eqnarray}
where $\rho$ is the fluid density, ${\bf u}$ is the velocity, $P$
the pressure, ${\bf j}$ is the electric current density, ${\rm B}$
is the magnetic field density and $e$ is the internal energy. In
Eq.\ (\ref{motion.eq}) ${\bf j}\times{\bf B}$ is the Lorentz force
giving rise to the Lorentz work ${\rm W}_{\rm L}$. ${\rm Q}_{\rm
v}$ and ${\rm Q}_{\rm J}$ are the viscous and Joule dissipation
respectively. ${\rm Q}_{\rm cool}$ is a Newtonian cooling term and
Re$_{\rm m}$ is the magnetic Reynolds number. ${\bf f}$ is an
external prescribed forcing term with an amplitude that is allowed
to evolve with time, keeping the average kinetic energy
approximately constant through both laminar and turbulent phases,
cf. \inlinecite{Archontis+ea03}.

Equations Eq.\ (\ref{mass.eq}-\ref{induction.eq}) are solved
numerically on a staggered mesh using derivatives and
interpolations that are of 6th and 5th order respectively in a
numerical scheme that conserves $\nabla\cdot{\bf B}=0$ exactly.
Numerical solutions are obtained on a grid of $128^{3}$ points,
the code by Nordlund and others e.g.\ \inlinecite{Nordlund+ea92}.

The chosen initial velocity field is similar to the classical ABC
flow \cite{Dorch2000} but does not contain cosine terms:
\begin{equation}
 {\bf u} = (\sin{z}, ~\sin{x}, ~\sin{y}), \label{nocosines}
\end{equation}
where the coordinates $(x, ~y, ~z)$ have a periodicity of ${2\pi}$
in all directions. The total kinetic helicity $\int_{\rm V} {\bf
u} \cdot \omega~ {\rm dV}$ (where $\omega = \nabla \times {\bf u}$
is the vorticity) is identically zero for this flow that has
previously been found to be a good candidate for fast dynamo
action: The growth rate of the magnetic field is known to increase
as a function of Re$_{\rm m}$ in the linear regime, at least until
Re$_{\rm m}=800$ \cite{Galloway+Proctor1992}. The initial magnetic
seed field is chosen to be weak and divergence-free with an
amplitude of $ 10^{-5}$ in non-dimensional units.

\begin{figure}
\centerline{\includegraphics[width=15pc]{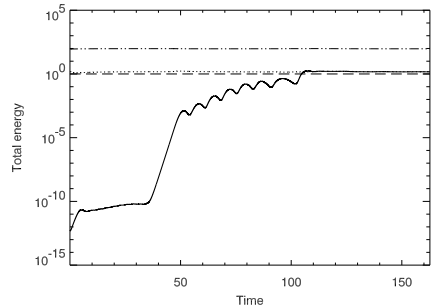}}
\caption{Total energies as a function of time for an experiment
with ${\rm Re}_{\rm m}=100$ and Re = 2. Here the solid curve is
magnetic energy ${\rm E}_{\rm mag}$, the dotted curve kinetic
energy ${\rm E}_{\rm kin}$ and the dotted-dashed curve the total
thermal energy ${\rm E}_{\rm th}$. The dashed horizontal line
indicates unity energy.} \label{fig1}
\end{figure}

\section{Results and discussion}
\label{results.sec}

We have performed several numerical experiments with varying
magnetic Reynolds number Re$_{\rm m}$ between 50--200 and kinetic
fluid Reynolds number Re between 2--450.

Initially during the linear, kinematic regime the total magnetic
energy ${\rm E}_{\rm mag}$ grows exponentially because of a small
difference between Lorentz work ${\rm W}_{\rm L}$ and the Joule
dissipation ${\rm Q}_{\rm J}$. Figure \ref{fig1} shows the
evolution of total energy in an experiment with ${\rm Re}_{\rm
m}=100$ and Re = 2: In this case the magnetic field was initially
give by a weak {\em uniform} seed field and this special topology
results in several linear dynamo modes being exited, but
eventually ${\rm E}_{\rm mag}$ saturates at equipartition within a
few fractions of a percent.

The balance between ${\rm W}_{\rm L}$ and ${\rm Q}_{\rm J}$ in the
linear regime does not hold in places with weak magnetic field and
little dissipation, as has also been shown in previous kinematic
dynamo experiments \cite{Archontis+ea03a}. A calculation, for low
${\rm Re}$, of the average net work $-{\rm W}_{\rm L} - {\rm
Q}_{\rm J}$ yields that most of the work responsible for the fast
dynamo action, occurs in places with less than 20\% of the maximum
dissipation and magnetic field strength and occupies 98\% of the
volume. There, the net work is 95\% of the maximum work level. In
the remaining 2\% of the volume the strong dissipation is balanced
by the work done on the field by the flow through advection and
convergence.

\begin{figure}
\tabcapfont
\centerline{%
\begin{tabular}{c@{\hspace{3pc}}c}
 \includegraphics[width=2in]{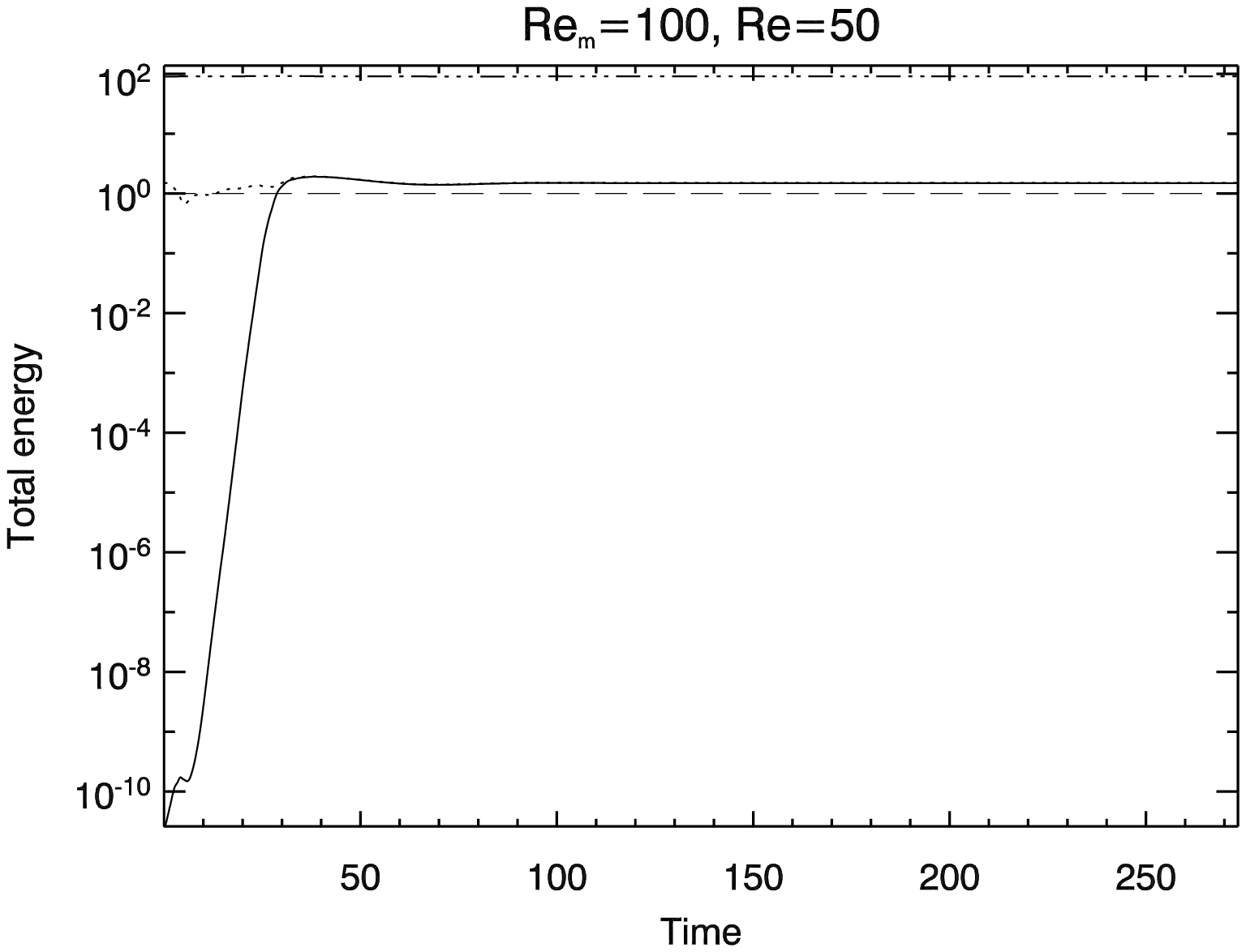} &
 \includegraphics[width=2in]{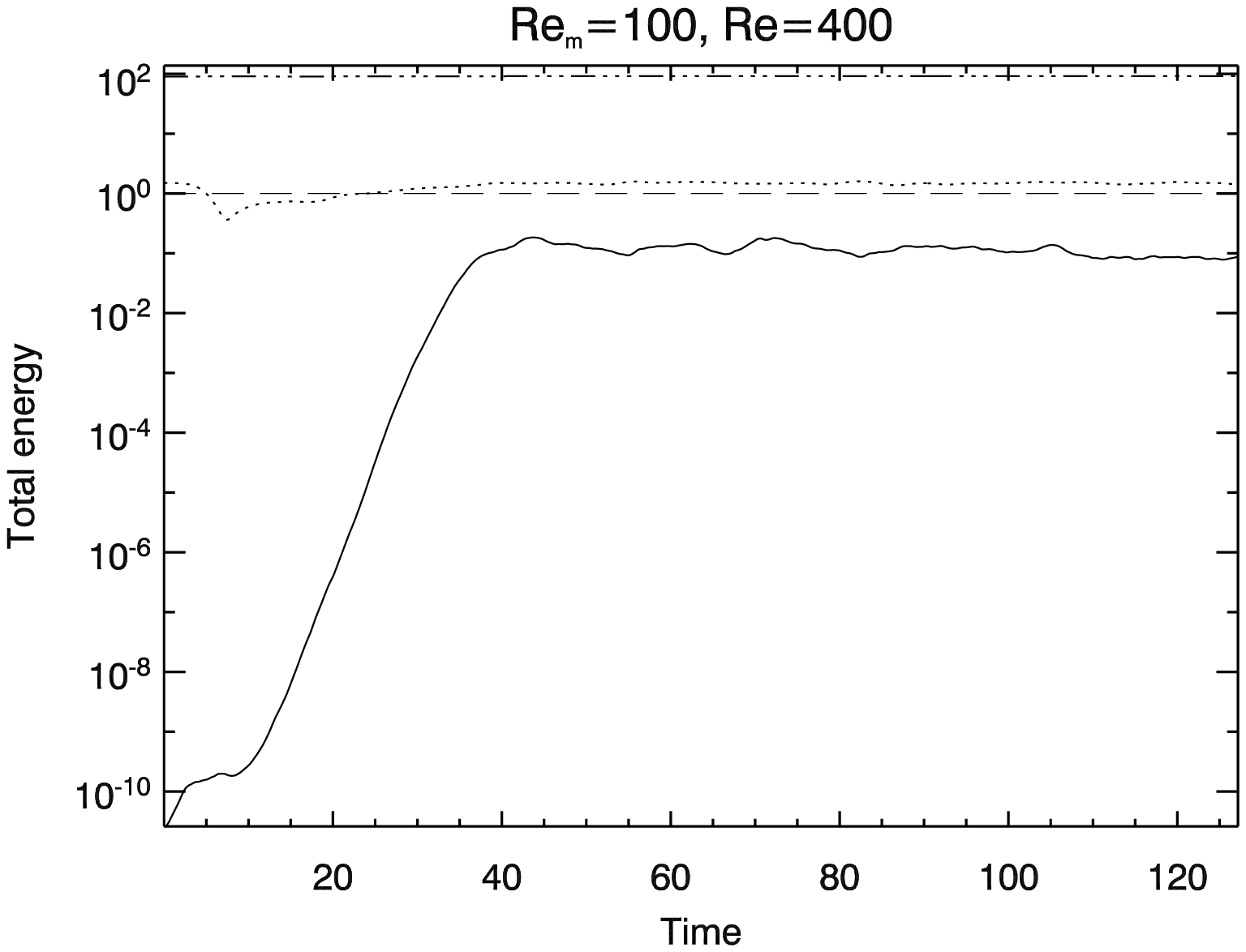}
\end{tabular}}
\caption{Total magnetic, kinetic and thermal energy as functions
of time for cases with ${\rm Re}_{\rm m}=100$ and Re = 50 (left)
and ${\rm Re}_{\rm m}=100$ and Re = 400 (right). Line styles are
the same as in Fig.\ \ref{fig1}.} \label{fig2}
\end{figure}

If the magnetic field is initialized by a {\em random} seed field
fewer linear modes appear before saturation: Figure \ref{fig2}
(left) shows magnetic energy as a function of time for a case with
${\rm Re}_{\rm m}=100$ and Re = 50 that also saturates near
equipartition, but only has one dominant, very fast growing mode
during the linear regime.

\begin{figure}
\tabcapfont
\centerline{%
\begin{tabular}{c@{\hspace{3pc}}c}
 \includegraphics[width=2in]{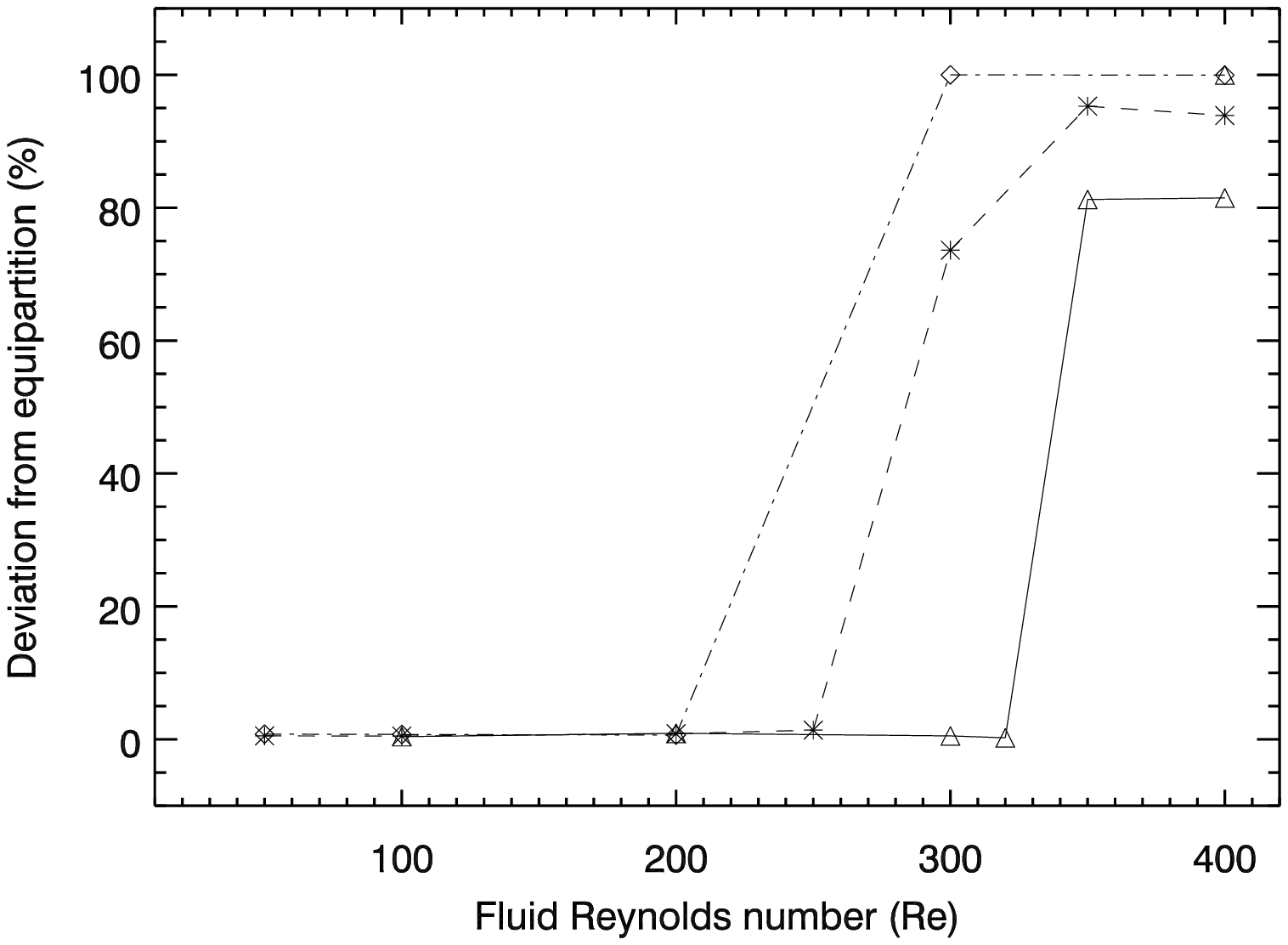} &
 \includegraphics[width=2in]{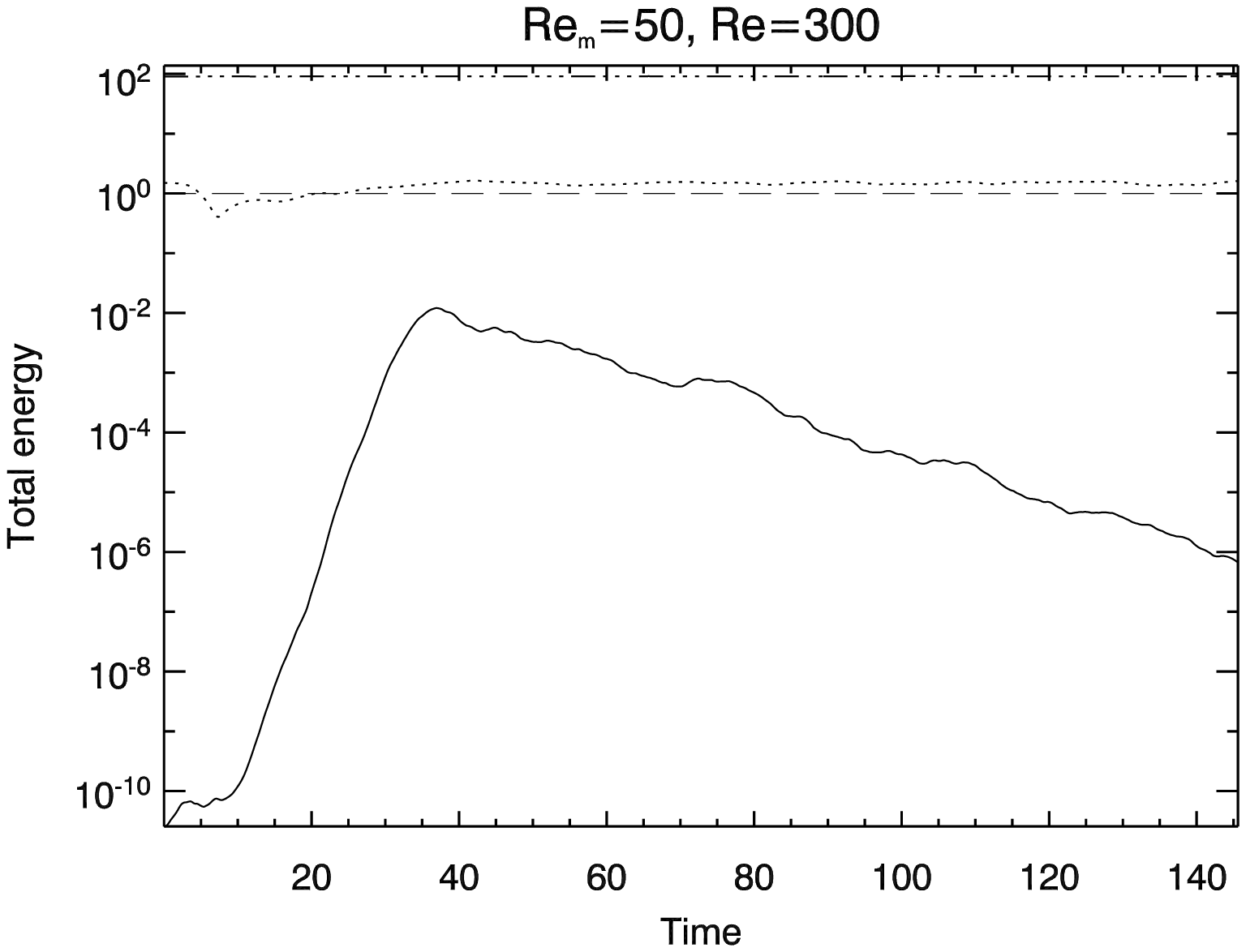}
\end{tabular}}
\caption{Left: A summary figure of the deviation (in percent) from
saturation at exact equipartition as a function of fluid Reynolds
number Re. Three regimes are shown corresponding to experiments
with ${\rm Re}_{\rm m}=200$ (full curve), ${\rm Re}_{\rm m}=100$
(dashed curve) and ${\rm Re}_{\rm m}=50$ (dashed-dotted curve).
Right: Total magnetic, kinetic and thermal energy as functions of
time for an experiment with ${\rm Re}_{\rm m}=50$ and Re = 300
corresponding to ${\rm Pr}_{\rm m}=1/6$ (right figure). Line
styles are the same as in Fig.\ \ref{fig1}. } \label{fig3}
\end{figure}

In these cases where Re is low, the flows are very laminar, as we
shall show below, but if Re is increased turbulence sets in. With
well-developed turbulence the magnetic energy saturates
significantly below equipartition as can be seen in Fig.\
\ref{fig2} (right) for a case also with ${\rm Re}_{\rm m}=100$ but
Re = 400, where in the non-linear regime ${\rm E}_{\rm mag}
\approx 0.2~ {\rm E}_{\rm kin}$. This dependence on Re is
summarized in Fig.\ \ref{fig3} (left) that shows the deviation
from exact equipartition, after saturation, as a function of Re
for three different values of ${\rm Re}_{\rm m}$: Apparently there
is a jump at Re $\sim 300$ below which the dynamo saturates at
exact equipartition. This scenario can also be quantified in terms
of magnetic Prandtl number ${\rm Pr}_{\rm m}$. There is a critical
value ${\rm Re}^{\rm (c)}$ below which the dynamo saturates below
equipartition and the value of ${\rm Re}^{\rm (c)}$ depends on
${\rm Pr}_{\rm m}$ roughly as ${\rm Re}^{\rm (c)} \sim 50~ {\rm
Pr}_{\rm m}^{0.5}$. At sufficiently low ${\rm Pr}_{\rm m}$ the
dominating dynamo mode is in fact a decaying mode, see e.g.\ Fig.\
\ref{fig3} (right) that shows the behavior of the total energy for
a case with ${\rm Re}_{\rm m}=50$ and Re = 300 corresponding to
${\rm Pr}_{\rm m} = 1/6$.

A tell-tale result is that the flow (for low ${\rm Re}$), after it
goes through a turbulent phase just prior to saturation, comes out
as a laminar solution with the magnetic and velocity field
parallel and equal in magnitude, see Fig.\ \ref{fig5} (left):
E.g.\ for the experiment with ${\rm Re}_{\rm m}=200$ and Re = 200
less than 2\% of the volume is occupied by field and flows
inclined by more than 10$^{\rm o}$. In this case, with unit
magnetic Prandtl number Pr$_{\rm m}$, the viscous and resistive
losses are both directly proportional to the velocity and magnetic
field strength, respectively, with the same constant of
proportionality and equal to $1/{\rm Re}$. For Pr$_{\rm m} \neq 1$
the two fields are still proportional but not exactly equal and
the saturation level deviates from exact equipartition in
fractions of a percent.

\begin{figure}
\tabcapfont
\centerline{%
\begin{tabular}{c@{\hspace{3pc}}c}
 \includegraphics[width=2in]{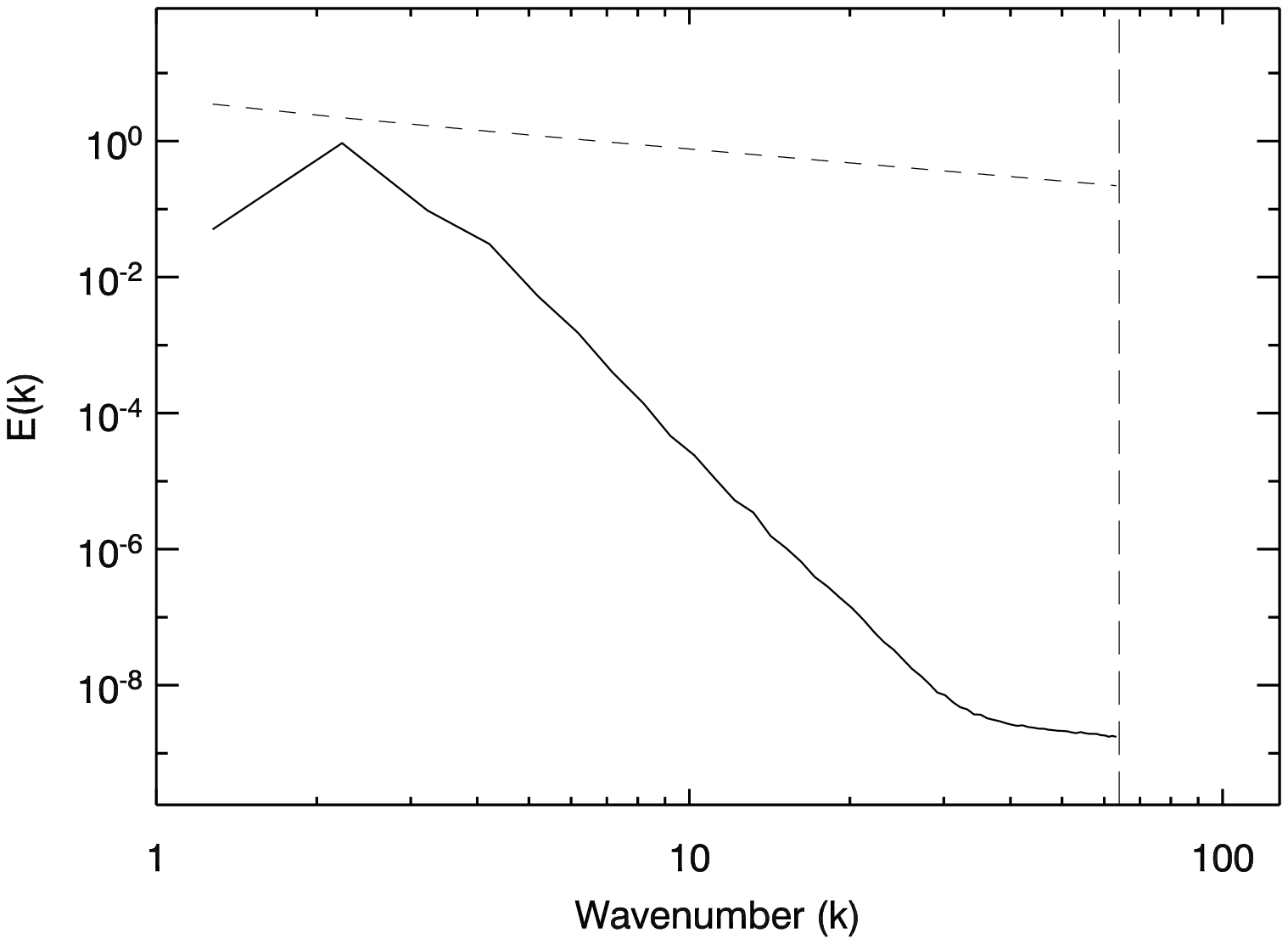} &
 \includegraphics[width=2in]{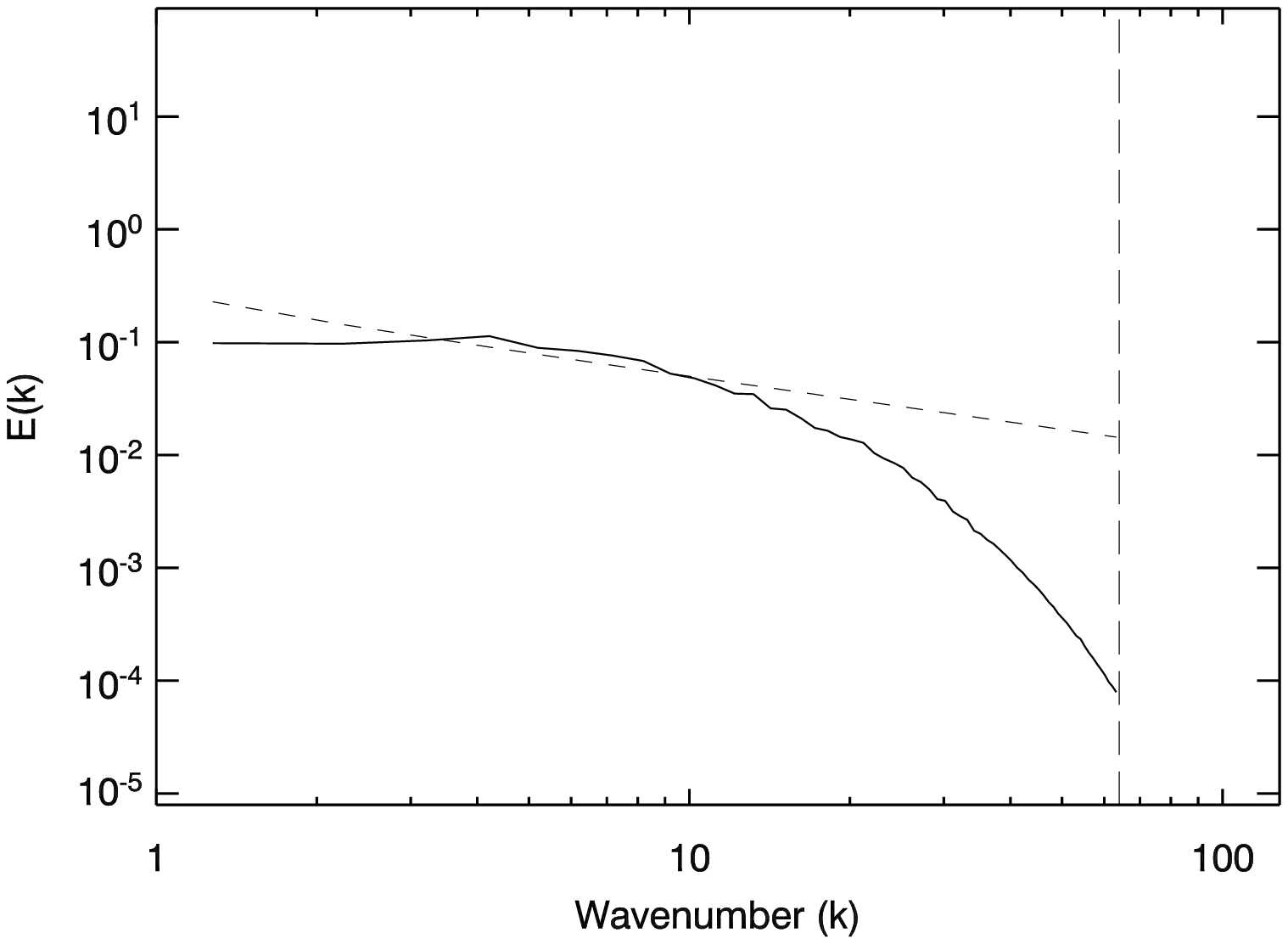}
\end{tabular}}
\caption{Energy spectra of magnetic energy as function of
wavenumber $k$ for for a case with ${\rm Re}_{\rm m}=50$ and Re =
200 (left), and a case with ${\rm Re}_{\rm m}=200$ and Re = 450
(right).} \label{fig4}
\end{figure}

In fact, there are marked differences between the structure of the
magnetic and velocity fields in the experiments as Re is
increased: This is true both in Fourier $k$-space as well as in
physical space. In $k$-space, for ${\rm Re}_{\rm m}=200$ and Re =
200, the magnetic and kinetic energy power are completely
coincident (with a ratio of $\sim 0.95$) except at the smallest
scales where there is slightly more power in the flow. The
magnetic energy has a maximum at $k \approx 2.2$ and the spectra
approximately follow a power-law with a power of -3, but for
$k\ge10$ become very steep, see Fig.\ \ref{fig4} (left). In cases
with ${\rm Re}_{\rm m}=200$ and higher Re there is also a lot of
power at large scales, but the spectrum is much flatter and
roughly follows Kolmogorov scaling for intermediate wavenumbers,
see Fig.\ \ref{fig4} (right). Examination of the magnetic energy
per wavenumber $k{\rm E}(k)$ reveals that most energy comes from
scales $k\approx10$ similar to what was found in models of
non-helical turbulence by \inlinecite{Haugen+ea03}.

In physical space, the behavior of the magnetic field and the
velocity flow can be quiet revealing. Hereafter, we consider first
the situation with Re $\le$ 200 and then a higher Re = 400 case.
During the exponential amplification of the magnetic energy, the
magnetic field has the form of strong sheets or tubes which are
spiraling around strong vortex tubes, cf. Fig.\ \ref{fig5} (left).
The weak background field is stretched by the flow and is folded
against the sheets in a highly twisted manner. In the saturation
regime the flow and the magnetic field have a less complex
configuration and consist of regions with stagnation points.
Figure \ref{fig5} (left) illustrates that there are two kinds of
flow stagnation points with spiraling streamlines in a plane
perpendicular to an axis of convergence or divergence,
respectively: Near the center of Fig.\ \ref{fig5} (left) almost on
a vertical line, it is illustrated how a stagnation point with a
spiral plane connects to a second stagnation point where the field
and stream lines form a plane perpendicular to an axis that in
turn approaches the spiral plane of the first stagnation point.
Magnetic null-points coincide with flow stagnation points, and
sites of maximum magnetic flow and field strength coincide; here
the magnetic and velocity fields take the form of collapsed
``blobs" of parallel field and stream lines that have been folded
and intensified. Next to these are sheets of strong vorticity.

\begin{figure}
\tabcapfont
\centerline{%
\begin{tabular}{c@{\hspace{3pc}}c}
 \includegraphics[width=2in]{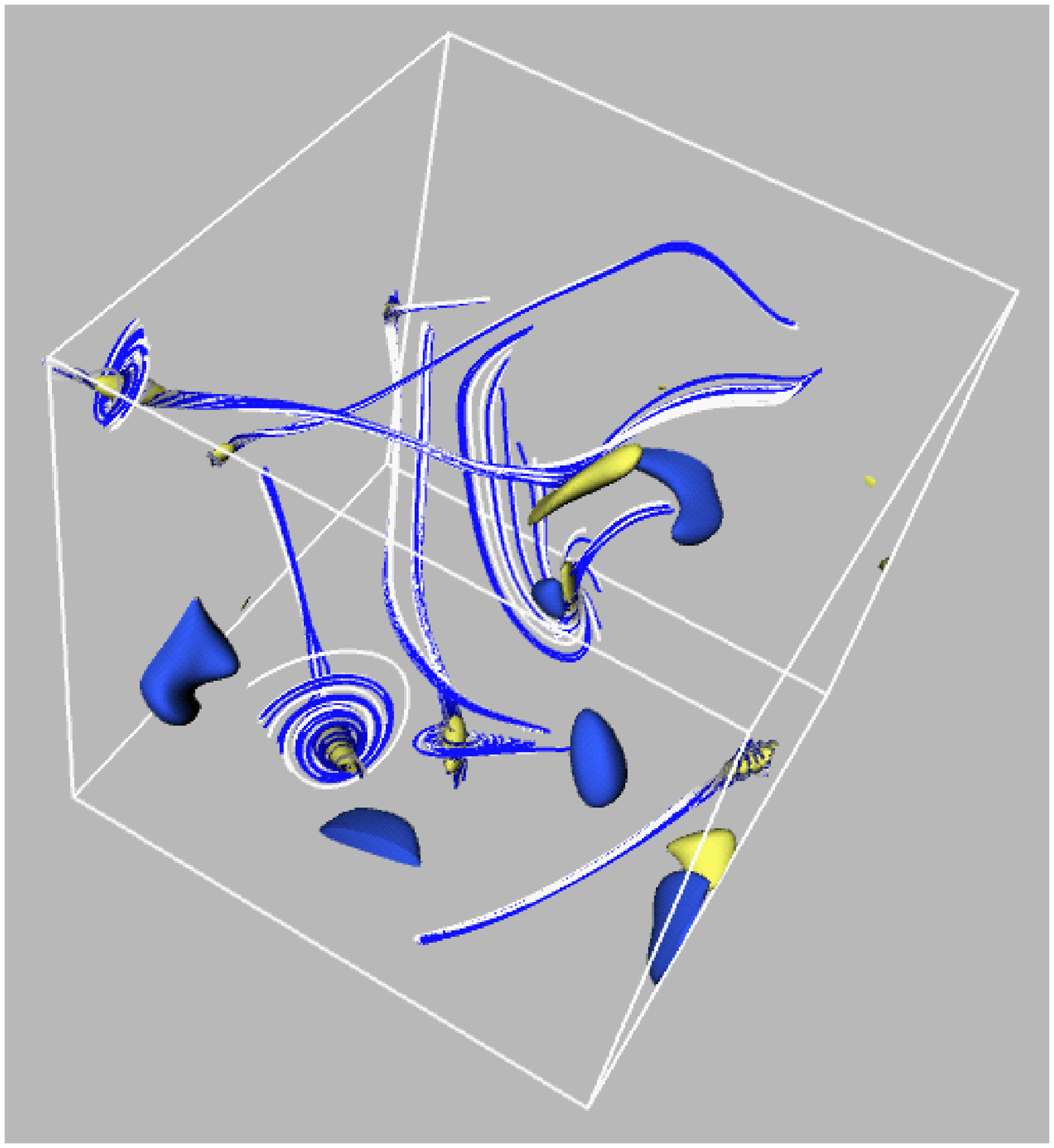} &
 \includegraphics[width=2in]{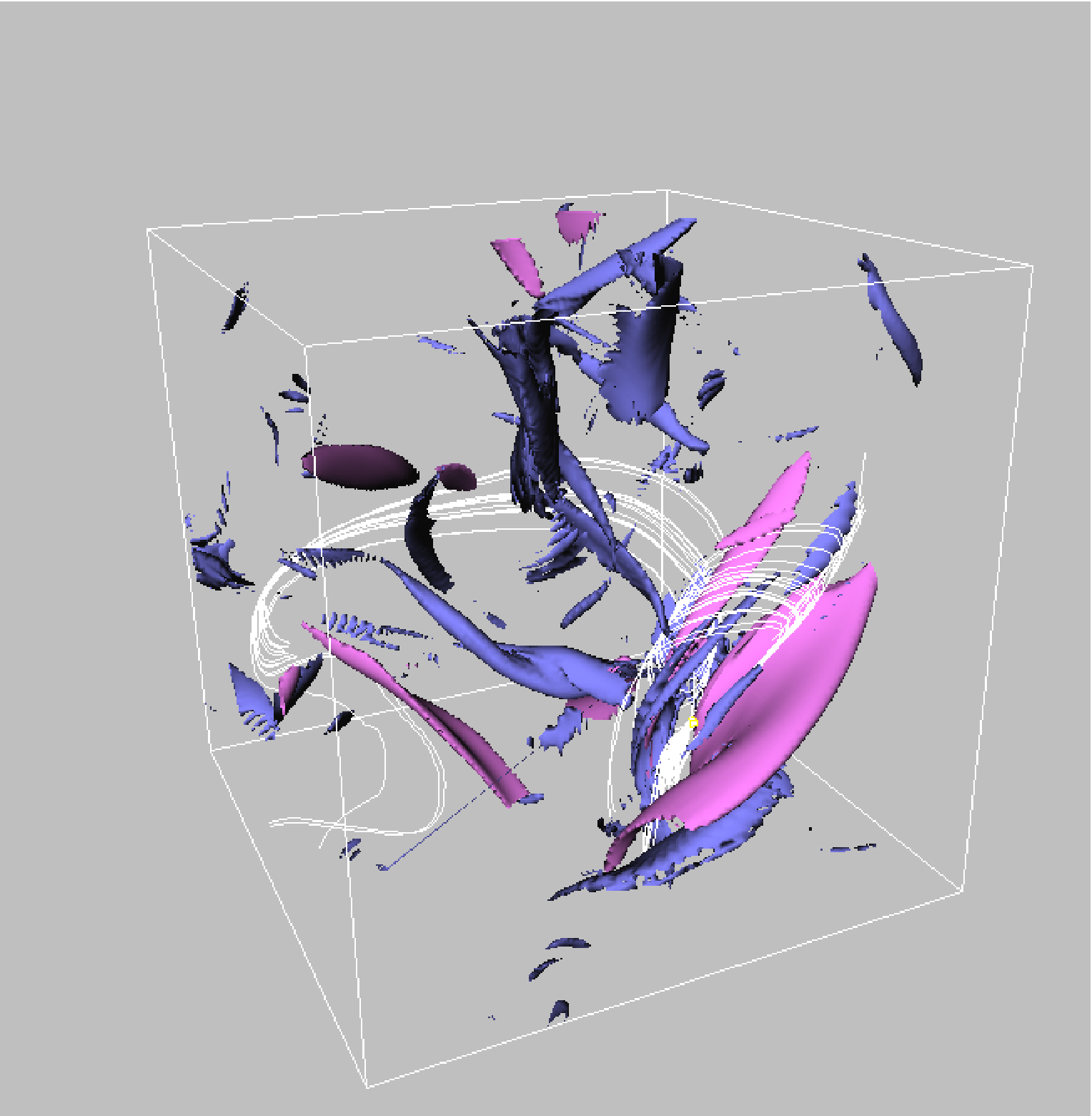}
\end{tabular}}
\caption{Left: Volume rendering of the magnetic field and velocity
flow in the experiment with ${\rm Re}_{\rm m}=200$ and Re = 200
during saturation. Isosurfaces of regions of low magnetic field
and flow (light structures) and local maxima (dark). The lines are
magnetic field lines (dark) and stream lines (light). Right:
Strong magnetic field (light) and vorticity structures (dark) in
the experiment ${\rm Re}_{\rm m}=200$ and Re = 400 during
saturation. Field lines traced from weak magnetic field (left side
of the box) are stretched and pile up in twisted sheets around
strong vortex tubes (to the right). } \label{fig5}
\end{figure}

In the case of the experiments with high Re that saturates below
equipartition, there are fluctuations in magnetic energy and
time-intervals where ${\rm E}_{\rm mag}$ increases or decreases
(Fig.\ \ref{fig2}, right). During the increase most of the work
occurs again in between the strong magnetic structures where
dissipation is low. The velocity has still a good grip on the
weakest part of the field and increases the magnetic energy by
stretching the field lines. The magnetic field has the form of
intermittent flux tubes or sheets that are twisted around regions
of high vorticity (Fig.\ \ref{fig5}, right). The twisting of the
magnetic field increases the tension of the magnetic field lines
and thereby the magnetic energy. Twisted sheets with like polarity
merge and those with opposite polarity reconnect, with the most
rapid release of energy occurring when ${\rm E}_{\rm mag}$ is
maximum. The reconnected field lines go back and replace the weak
field which was initially stretched by the flow. When the magnetic
energy decreases the field lines start to untwist and magnetic
structures are split into smaller pieces. This mode is reminiscent
of the kinematic ABC-flow dynamo that also exhibits an undulatory
behavior of ${\rm E}_{\rm mag}$ as magnetic field lines are first
experiencing increased tension boosting ${\rm E}_{\rm mag}$,
subsequently leading to reconnection decreasing ${\rm E}_{\rm
mag}$ but replenishing the supply of weak field for the flow to do
Lorentz work on \cite{Dorch2000}. In this non-linear case however,
there is no periodic behavior and there is no symmetry to the
flow, in the form of e.g.\ stagnation points and hence the
reconnected weak magnetic field lines are not twisted to form new
flux tubes by the same velocity structures that formed their
``ancestors".

\section{Conclusion}
\label{conclusion.sec}

We have presented results from new numerical experiments with
non-linear dynamo action by a non-helical flow while varying
characteristic dimensionless parameters: No matter whether in the
linear or saturated regimes, or whether the flow is laminar or
turbulent, dynamo action occurs primarily where the field is weak.
3-d visualization of the flow and the magnetic field shows similar
structures associated with stretching, folding and subsequent
amplification of the magnetic energy, in both the linear and
non-linear equilibrated regime. Additionally, we find that there
is a critical fluid Reynolds number above which the dynamo no
longer saturates at or close to equipartition, but at a lower
level characteristic of turbulent astrophysical dynamos. In terms
of energy spectra there are marked differences between the high
and low Re cases, but in physical space we identify some
individual common processes that are likely to be a major part of
the dynamo mechanism. Our intention was to illustrate that
studying dynamo action by these kind of flows is a viable way to
understand astrophysical dynamos. In a forthcoming paper we will
follow up with a broad analysis in physical space of the numerical
experiments presented here: We believe that the key to
understanding the nature of non-linear dynamo action is to answer
such questions as ``where do the magnetic field lines come from,
and where do they go?"

\acknowledgements

SBFD was supported by the Danish Natural Science Research Council.
Access to computational resources granted by the Danish Center for
Scientific Computing is acknowledged.

\end{article}
\end{document}